\documentclass{article}
\usepackage{setspace}
\def\be{\begin{equation}}
\def\ee{\end{equation}}
\def\o{\Omega_{b}}
\def\bq{\begin{eqnarray}}
\def\eq{\end{eqnarray}}
\def\rdr{\frac{\dot{a}}{a}}
\def\rdrsq{\frac{\dot{a^2}}{a^2}}
\def\bdbsq{\frac{\dot{b^2}}{b^2}}
\def\bdb{\frac{\dot{b}}{b}}
\def\rddr{\frac{\ddot{a}}{a}}
\def\bddb{\frac{\ddot{b}}{b}}
\def\l{\Lambda}

\begin{document}
\renewcommand{\do}{\mbox{$\partial$}}
\title{On brane cosmological solutions} 
\author{H.~K.~Jassal \thanks{E-Mail: hkj@iucaa.ernet.in} \\
{\em Inter University Centre for Astronomy and Astrophysics,
\thanks{Present address}} \\  
{\em Post Bag 4, Ganeshkhind, Pune-411 007, India.} \\
{\em and} \\
{\em Harish-Chandra Research Institute,} \\
{\em Chhatnag Road, Jhusi, Allahabad-211 019, India.} \\ 
}
\maketitle
\begin{center}
\Large Abstract
\end{center}
\large
This paper studies some cosmological consequences of the five
dimensional, two brane Randall-Sundrum brane scenario.
The radius of the compact extra dimension is taken to be time dependent.
It is shown that the cosmology consistent with the two brane
Randall-Sundrum model is a power law expansion of the universe, with
scale factor growing as $t^{1/2}$.
The two branes tend to move towards each other with time.
Some comments are made on the contribution of surface terms in
deriving the four dimensional effective action.  
\\

PACS number(s) : 04.50.+h, 95.30.SfA

\pagebreak

The idea that we may be living in dimensions higher than four has been 
under investigation for several decades now.
In fact, the original Kaluza-Klein theory  is as
old as Einstein's theory of gravity.
Generically, Kaluza-Klein theories (for a review see,
\cite{overduin}) are theories of gravitation formulated in higher
dimensions, some of which are compactified.
The possibility that there are additional dimensions which remain
small was put forward as a means of unifying the electromagnetic and
gravitational fields as components of a single higher-dimensional
field. 
The matter fields arise out of geometry and the theory essentially is a 
minimal extension of Einstein's theory of relativity to higher
dimensions.

Recently there have been attempts to address the mass hierarchy
problem by invoking the presence of extra dimensions. 
One assumes localisation of matter fields on a
four-dimensional brane embedded in a $D=4+n$ dimensional bulk
\cite{dvali1,dvali2}, while
gravity can propagate in all the $D$ dimensions. 
In these higher dimensional models, it is assumed that the geometry
is a direct product of the four dimensional spacetime with an
$n$-dimensional compact manifold.
From the point of view of an observer on the four-dimensional brane,
the Planck scale is given by
$M_{Pl}^{2}=M^{n+2}V_{n}$, where $V_{n}$ is the compact space volume
and $M$ is the fundamental Planck scale.
If the radius of the extra dimension $R_{n}$ is large, i.e. of the
order of a millimeter, the fundamental Planck scale can be of the
order of a TeV. 
The phenomenology which arises as a result is  expected to be tested in
the next generation of particle colliders.
However the demand that the extra dimensions be large introduces an
implicit hierarchy  and is therefore not a satisfactory 
solution to the problem.

Randall and Sundrum (RS) \cite{randall1} proposed a five-dimensional
model based on a non-factorisable geometry.
The ``warp'' factor, which scales the four dimensional spacetime with
respect to the extra dimension, is a rapidly changing function of the
extra dimension.
This obviates the need that extra dimensions be large.
The hierarchy between the four dimensional Planck scale and the
fundamental scale of the theory is resolved because of the presence of 
the exponential ``warp'' factor.
The first RS model consists of  two four dimensional 
branes which are defects in a five dimensional anti-deSitter
background.   
One is a positive tension Planck brane and the other is the brane on
which standard model particles are confined, which has a negative tension
and is called the TeV brane. 
A variant of this higher dimensional scenario is the one brane world,
in which the Planck brane is taken  to infinity \cite{randall2} and
one has a four-dimensional brane embedded in a five-dimensional
anti-deSitter bulk.  

Since this new class of models have non-factorisable warped geometry, 
they are fundamentally different from the usual Kaluza-Klein theories,
and hence  involve different phenomenological issues.
One of the reasonable questions to ask is how these  will
affect the early universe cosmology.
In fact, a large number of papers (for instance
\cite{cosmo1,cosmo2,cosmo3,cosmo4}) have investigated the cosmological
aspects of the  scenarios proposed by Arkani-Hamed et al. and by
Randall and Sundrum.  
The motivation for such studies is to see if the brane scenario can
provide a unified solution to the mass hierarchy problem as well as
some problems in cosmology such as inflation.

Motivated by the above arguments, we study some implications of
five-dimensional warped geometry on early universe cosmology.
The spacetime structure considered here is the  RS two-brane
model. 
We first briefly review the model under consideration. 
It is shown that on integrating out the extra dimension from the
action, we have an extra term not dealt with so far in papers.
The main result in this paper is the universe has a power law
expansion. 
Here we have taken radion dynamics into account; radion being the
field associated with the fluctuations of the extra dimension.
There is no inflation and the branes tend to move
towards each other with time.

The five-dimensional spacetime is a slice of anti-deSitter geometry,
where we have a negative cosmological constant.
Two 3-branes  are located at fixed points of orbifold $S^{1}/Z_{2}$.
In other words, the extra fifth dimension is a circle with opposite
points identified.
Following \cite{csaki} we take the two orbifold points to be situated
at $y=0$ and $y=1/2$.
The positive tension Planck brane is located at $y=0$ and the
negative tension TeV brane is situated at $y=1/2$.

The five dimensional action for anti-deSitter spacetime is given by 
\bq
\label{eq:fiveaction}
S &=& 2 \int d^4x \int_{0}^{1/2} dy \sqrt{-G} \left(M^{3} R-\l \right)
\\ \nonumber
&+&\int d^4x \sqrt{-g^{(+)}} \left(L^{+}-V^{+} \right)  \\ \nonumber
&+&\int d^4x \sqrt{-g^{(-)}} \left(L^{-}-V^{-} \right)
\eq
where 
\be
V^{+}=-V^{-}=12 m_{0} M^{3},~~\l=-12 m_{0}^{2} M^{3},
\ee
and where $R$ is the five dimensional Ricci scalar, the bulk
cosmological constant is denoted by $\l$ and $M$ is the five 
dimensional Planck mass. The constant $m_0$
is same as the parameter $k$ in \cite{randall1,randall2}. 
The $(+)$ sign denotes the Planck brane and $(-)$ sign represents the
negative tension TeV brane.
The matter fields on the positive and negative tension branes are
$L^{+}$ and $L^{-}$ respectively, while $V^{\pm}$ represent the brane
tensions on positive and negative tension branes respectively.

The metrics on the two four-dimensional branes are therefore given by
\bq
g_{\mu \nu}^{(+)} = G_{\mu \nu}(x^{\mu}, y=0)~~{\rm and} \\ \nonumber
g_{\mu \nu}^{(-)} = G_{\mu \nu}(x^{\mu},y=1/2)
\eq
The five-dimensional Einstein equations of
motion are solved by the metric
\be
ds^2=e^{-2 m_0 r_c \mid y \mid} \eta_{\mu \nu} dx^{\mu} dx^{\nu} +
r_{c}^{2} dy^{2}
\ee
where $\eta_{\mu \nu}$ represents the flat four dimensional 3-brane
while $r_c$ is the radius of the extra dimension.

The relation between the four dimensional Planck scale $M_{Pl}$ and the
fundamental scale $M$ in the theory
$$
M_{Pl}^2=\frac{M^3}{m_0}\left[1-e^{-2m_0r_c}\right]
$$
Because of the exponential warp factor, to scale the hierarchy required between
the two scales, we need the product $m_0 r_c$ to be of the order of 50.
Therefore this provides us with an interesting approach to solve the
mass hierarchy problem.  

To accommodate cosmological solutions, we deviate from the
RS flat branes geometry and assume the modulus $r_c$ to be time
dependent \cite{csaki,debchou}. 
The five-dimensional metric ansatz is \cite{csaki} 
\be
ds^2=e^{-2 m_{0} b(t) \mid y \mid} g_{\mu \nu} dx^{\mu} dx^{\nu} +
b^{2}(t) dy^{2}
\ee
with the four-dimensional spacetime being described by the spatially
flat Friedmann-Robertson-Walker metric
\be
g_{\mu \nu}={\rm diag} (-1, a^2(t), a^2(t), a^2(t))
\ee
where $a(t)$ is the scale factor.
We take the same notation as in Ref. \cite{csaki}.

Using the above metric ansatz, we integrate over the extra dimension
from the action given in Eq. (\ref{eq:fiveaction}).
The four dimensional action can be written as
\bq
\label{eq:4dact}
S_{eff}&=&-\frac{3}{k^2 m_0} \int d^4 x \left[ \left(1-\o^2\right)
\rdrsq \right. \\ \nonumber
&+& \left.m_0 \o^2 \rdr \dot{b}-\frac{1}{4}m_0^2 \o^2 \dot{b^2}\right] \\
\nonumber
&-&
\frac{1}{k^2}\int d^4 x a^3 8 m_0 \left(1-e^{-2m_0b} \right)
\eq
where $\o=e^{-m_0 b(t)/2}$ and $k^2=1/2M^3$.

The action can further be written as
\bq
\label{eq:4dact1}
S_{eff}&=&-\frac{1}{2 k^2 m_0}\int d^4x a^3(t) \left[(1-\o^{2} ) R_{4}
\right. \\ \nonumber
&-& \left.\frac{3}{2} m_{0}^2 \o^{2} \dot{b}^2 + 16 m_0^2 \left(1-e^{-2m_0b} \right) \right]
\eq
The four dimensional Ricci scalar is denoted by $R_4$.
The last `extra' term in the above action gets canceled if one
includes surface terms in the action. 
A discussion regarding cancellation of these terms is presented
 in Appendix A.

The cosmological equations of motion  obtained from the action without
the extra term are given by 
\bq
3H^2&=&\frac{3 \dot{\o}^2}{1-\o^2} 
+6H \frac{\o \dot{\o}}{1-\o^2} \\ \nonumber
2\dot{H}+3H^2&=&-\frac{\dot{\o}^2}{1-\o^2}+4H \frac{\o
\dot{\o}}{1-\o^2} \\ \nonumber
&+& \frac{2 \o \ddot{\o}}{1-\o^2} \\ \nonumber
6 \frac{\ddot{\o}}{\o}&+&18H\frac{\dot{\o}}{\o}=0
\eq
which have solutions given by 
\bq
H&=&\frac{1}{2t} \\ \nonumber
\o(t)&=&\pm 1+At^{-1/2}
\eq
where $H=\frac{\dot{a}(t)}{a(t)}$, $\o(t)=e^{-m_0 b(t)/2}$ and $A$ is a  positive integration constant.
Since $\o$ is an exponential function, we need to consider only the
$+$ branch.

Here, $\o$ approaches $1$ as time $t$ tends to a large value.
In this limit $b(t)$ approaches 0, i.e. the two branes are
moving closer and closer to each other. 
If one considers the matter on the brane to be radiation only, one can 
again show (with a similar calculation) that the power law expansion
as $a(t) \sim t^{1/2}$ holds.

A mechanism to stabilise the radius of the extra dimension to its
equilibrium value was proposed in Ref. \cite{gw}.
Here one considers a bulk scalar field which is allowed to
propagate in all the five dimensions.
The scalar field,  in general, has different vacuum expectation values
on the two branes.
The gradient created along the extra dimension and the potential
energies, lead to a radion potential.
This potential has the properties required for inflation.
The scalar field rolls over and settles at its minimum hence
stabilising the value of $b(t)$.
This brings us to an interesting question if the scalar field 
remains at its minimum and does not oscillate out of the potential
well (work in progress and will be reported elsewhere) \cite{hkj}.

This paper presents some cosmological solutions allowed by the five
dimensional Randall-Sundrum two brane scenario.
The solutions to cosmological equations of motion, reveal that even if
the extra term  is absent, one obtains a power law cosmology.
However, the branes tend to move towards each other with time, leading 
to a future collapse.
The situation remains the same if one assumes the presence of radiation 
on the brane. 
Inflationary solutions can be obtained only if one assumes the
presence of an external potential. 
Interestingly, if one makes a variable transformation as in
\cite{debchou} the solutions are still power law in the new time
parameter.
It is conjectured by many earlier authors that the potential suggested by
Goldberger and Wise \cite{cline} to stabilise the radion to its
equilibrium value can be the potential driving inflation. 
The equations of motion then have to be solved numerically to
investigate the behaviour of the cosmological equations. 
Work on this is in progress.
We integrate out the extra dimension to obtain the four-dimensional
effective action. 
The effective action obtained from such an exercise  contains an
extra term which has not been considered in papers so far. 
It is conjectured that this term arises because the surface terms 
have not been included in the calculations.
Another possible explanation could be because of metric ansatz being
put at the action level, which  gives different results from when it
is put in the equations of motion. 
Hopefully the underlying connection will become clear with further
work.

\section*{Acknowledgements}
The author is thankful to S. Naik and D. Jatkar for
useful discussions during the initial stages of this work. 
Thanks are due to P. Majumdar, A. Mukherjee, T. Padmanabhan and
K. Subramanian for for giving comments on the manuscript. 
The major part of this work was done while the author was at
Harish-Chandra Research Institute.

\appendix
\section{Calculation of higher dimensional Ricci tensor and the
surface term}

In this appendix we present the steps in the calculation of the
four-dimensional action.
The five dimensional Ricci scalar is
\bq
R&=&-20 m_0^2 - e^{2 m_0 b(t)y}\left\{-6 \rdrsq \right. \\ \nonumber
&-& \left. 6\rdr \bdb -6 m_0^2 y^2
\dot{b}^2 
-6 \rddr -2 \bddb \right\} \\ \nonumber
&-& y  e^{2 m_0 b(t)y} \left\{ 18
m_0 \rdr \dot{b} +4 m_0 \bdbsq + 6 m_0 \ddot{b} \right\}
\eq

We consider the first term in the five dimensional action, {\it viz.}
$$
S_1=2 \int dx^4 \int_{0}^{1/2} \sqrt{-G} \left(M^{3} R-\l \right)
$$
The constant term $-20m_0^2$ combines with $\l=12 m_0^2 M^3$ and the
first term in the five dimensional action contains 
\be
S_1=2 M^3 \int d^4 x \int_0^{1/2} -e^{-4 m_0 b y}a^3(t) b 8 m_0^2
\ee
in addition to the other terms mentioned above.

Integration over the extra dimension gives 
\be
S_1(eff)=-2 M^3 \int d^4x~ a^3(t) 2m_0 \left[ 1-e^{-2 m_0 b(t)}\right]
\ee
Combining the terms containing $V^{(+)}$ and  $V^{(-)}$, we are left
with a term 
\bq
S_{(eff)}&=&-2 M^3 \int d^4 x  a^3(t) 8m_0 \left[ 1-e^{-2 m_0
b(t)}\right] \\ \nonumber
&+&~~\mathrm{other~terms} 
\eq

In general, the Ricci scalar in $d+1$ dimensions  contains a term
proportional to $d(d+1)$ and the cosmological constant has a term
proportional to $d(d-1)$.  
This leaves a quantity proportional to $2d^2$ on combining the above
two.
Hence there will remain an extra factor in $d$ dimensions, and if
present may have  nontrivial contribution to the solutions.

We now come to computation of the surface term.
In the original Randall-Sundrum brane world, the metric is
\be
ds^2=e^{-2 k y} g_{\mu \nu}dx^{\mu}dx^{\nu}+dy^2.
\ee
The Ricci tensor and scalar in this case are
\be
R_{\mu \nu}=-4k^2 g_{\mu \nu}~~~~\mathrm{and}~~~~R=-20k^2
\ee
The term $\frac{\delta R_{\mu \nu}}{\delta g^{\mu \nu}}$ is $-8 k^2$
the number required for cancellation! 
The factor is same for the time dependent case considered above.

Since the five dimensional spacetime is finite, the surface terms
(which do not contribute for an infinite spacetime) should
contribute in this case.
Therefore, we have to take into account the contribution from the term
$\frac{\delta R_{\mu \nu}}{\delta g^{\mu \nu}}$ while calculating the
equations of motion.
Indeed, for the `static' Randall-Sundrum case this term cancels the `extra'
term.
One requires this term in the action (to cancel away the surface
term) and integration along $y$ leads to the equations of motion
as in other papers.
The same factor is required for the time dependent case too.

An alternate explanation is that we have
assumed the form of metric at the action level. 
The equations of motion obtained from this transformed action, do not
give a general solution. 
These solutions are a subset of those obtained by solving the
equations of motion obtained from the full action.
Therefore, a more general calculation does away with this term.
Comparing our results with another paper dealing with similar
aspects, inflationary solutions to the equations of
motion are obtained in \cite{debchou}. 
Incidentally, the potential used in the paper is same as
what one would obtain if one makes the same field transformation in the
`extra' term.


\end{document}